\documentclass[eqsecnum,aps,preprint]{revtex4}
\usepackage{epsfig,graphicx}
\usepackage[fleqn]{amsmath}
\usepackage{amssymb}
\usepackage{nicefrac}
\usepackage{dsfont}
\usepackage{booktabs}

\newcommand{\Tr}{\text{Tr}}
\newcommand{\TrCD}{\text{Tr}_\text{C,D}}
\newcommand{\TrC}{\text{Tr}_\text{C}}
\newcommand{\TrD}{\text{Tr}_\text{D}}
\newcommand{\ld}{: \hspace{-4pt}}
\newcommand{\rd}{\hspace{-4pt} :}

\newcommand{\D}[1]{\mathrm{d} \hspace{-1pt}#1}
\newcommand{\Dp}[2]{\mathrm{d}^{#1} \hspace{-1pt}#2}

\DeclareSymbolFont{rsfs}{U}{rsfs}{m}{n}
\DeclareSymbolFontAlphabet{\mathrsfs}{rsfs}

\begin{document}

\title{Chiral QCD sum rules for open charm mesons}

\author{\sc T.\ Hilger${}^{1)}$, 
B.\ K\"ampfer${}^{1,2)}$,
S. Leupold${}^{3)}$}

\affiliation{
${}^{1)}$Forschungszentrum Dresden-Rossendorf, PF 510119, D-01314 Dresden, Germany\\
${}^{2)}$TU Dresden, Institut f\"ur Theoretische Physik, 01062 Dresden, Germany\\
${}^{3)}$Institutionen f\"or fysik och astronomi, Uppsala Universitet, Sweden}

\begin{abstract}
QCD sum rules for chiral partners in the open-charm meson sector are presented
at nonzero baryon net density or temperature.
We focus on the differences between pseudo-scalar and scalar
as well as vector and axial-vector $D$ mesons 
and derive the corresponding Weinberg type sum rules.
This allows for the identification of such QCD condensates which drive the non-degeneracy 
of chiral partners in lowest order 
of the strong coupling $\alpha_s$ and
which therefore may serve as ``order parameters'' 
for chiral restoration (or elements thereof).
\end{abstract}

\maketitle

\section{Introduction} \label{sct:introduction}

Chiral symmetry and its breaking pattern represent important features 
of strong interaction physics.
The non-degeneracy of chiral partners of hadrons is considered to be a direct hint 
to the spontaneous chiral symmetry breaking in nature which characterizes the QCD vacuum.
In fact, the distinct difference of iso-vector--vector and iso-vector--axial-vector 
spectral functions deduced from $\tau$ decays 
\cite{Schael:2005am,Ackerstaff:1998yj} gives one of the empirical and precise evidences 
for the breaking of chiral symmetry.
The low-energy strengths of the mentioned spectral functions, 
concentrated in the resonances $\rho(770)$ and $a_1(1260)$, 
deviate strongly from each other and from perturbative QCD predictions.
(For a dynamical interpretation of the two spectra see \cite{Wagner:2008gz,Leupold:2009nv}.)
This exposes clearly the strong non-perturbative effects 
governing the low-energy part of the hadron spectrum.
The spontaneous symmetry breaking is quantified by the chiral condensate
$\langle \bar q q \rangle$, which plays an important role in the
Gell-Mann--Oakes--Renner relation connecting hadronic quantities and
quark degrees of freedom (cf.~\cite{GellMann:1968rz,Colangelo:2001sp}).

Weinberg type 
chiral sum rules for differences of moments between light vector and axial-vector 
spectral functions have been developed for vacuum \cite{Weinberg:1967kj,Das:1967zz} 
and for a strongly interacting medium at finite temperature \cite{Kapusta:1993hq}.
It is the aim of the present paper to present analog sum rules for the scalar and 
pseudo-scalar as well as vector and axial-vector
mesons composed essentially of a heavy and a light quark, 
e.g.\ $u \bar c$ and $d \bar c$ realized in vacuum as $D^\pm(1867)$, $J^P = 0^-$, $D^0(1865)$, $J^P = 0^-$, $D^\ast_0(2400)^{0}$, $J^P = 0^+$, $D^\ast(2007)^0$, $J^P = 1^-$, $D^\ast(2010)^\pm$, $J^P = 1^-$, $D_1(2420)^0$, $J^P = 1^+$ (there is no confirmed charged scalar or axial-vector state in the open charm sector) \cite{Nakamura:2010zzi}.
Such open charm degrees of freedom will be addressed in near future by 
the CBM \cite{CBM} and PANDA \cite{PANDA} collaborations at FAIR
in proton-nucleus and anti-proton-nucleus reactions \cite{Friman:2011zz}.
Accordingly we are going to analyze the chiral sum rules in 
nuclear matter.

Chiral symmetry is explicitly broken by nonzero quark masses, but quantum chromodynamics is still approximately invariant under transformations which are restricted to the light quark sector.
The associated conserved currents are the well known light quark vector and axial-vector currents.
Their mixing under chiral transformation and Wigner's non-degeneracy of the hadron spectrum is interpreted as the spontaneous breakdown of chiral symmetry.
Although currents, related to mesons which are represented in the quark model as composed of a light quark and a heavy quark, are neither conserved nor associated with a symmetry transformation, vector and axial-vector currents still mix under a transformation which is restricted to the light quark sector.
Consider for example the infinitesimal rotation in flavor space
$\psi \rightarrow e^{-i t^a \Theta^a \Gamma} \psi \approx \left(1-i t^a \Theta^a \Gamma\right)\psi$,
where $t^a \in {\rm SU}(n_f)$, $\Gamma = \mathds{1}$ for the vector transformations and $\Gamma = \gamma_5$ for the axial transformations, and $\Theta^a$ denote a set of infinitesimal rotation parameters (rotation angles).
The vector current transforms as
\begin{equation}
	j_\mu^{\rm V, \tau} = \bar \psi \gamma_\mu \tau \psi
	\rightarrow j_\mu^{\rm V, \tau} \pm i \bar \psi \Gamma \gamma_\mu \Theta^a \left[ t^a , \tau \right]_- \psi
	\: ,
\end{equation}
where $+$ $(-)$ refers to the vector (axial) transformations and $\tau$ denotes a matrix in flavor space.
In the 3-flavor sector $\psi = (\psi_1, \psi_2, \psi_3)$, where the first two quarks are light, the choice $\vec \Theta = (\Theta_1, \Theta_2, \Theta_3,0,\ldots,0)$ and $2 t^a = \lambda^a$, the well-known Gell-Mann matrices, clearly leaves the QCD Lagrangian invariant.
To be specific, let us consider $\tau = (\lambda^4 + i\lambda^5)/2$, where the corresponding current transforms as
\begin{equation}
	j_\mu^{\rm V, \tau} = \bar \psi_1 \gamma_\mu \psi_3
	\rightarrow j_\mu^{\rm V, \tau \prime} = j_\mu^{\rm V, \tau} - \frac i2 \left( \bar \psi_1 \Gamma \gamma_\mu \psi_3 \Theta_3
	+ \left(\Theta_1+i\Theta_2\right) \bar \psi_2 \Gamma \gamma_\mu \psi_3
	\right)
	\: .
\end{equation}
Obviously, the heavy-light vector current mixes with heavy-light axial-vector currents if an axial transformation, with $\Gamma = \gamma_5$, is applied.
Analog expressions hold for spin-0 mesons, and the result is the same as in the spin-1 case, but with the replacements $\gamma_\mu \to \mathds{1}$ for the vector transformation and $i \gamma_\mu \left[ t^a , \tau \right]_- \to \left[ t^a , \tau \right]_+$ for the axial transformation.
Hence, a symmetry and its spontaneous breakdown in the light quark sector must also be reflected in the spectrum of mesons composed of a heavy and a light quark.
In particular, the splitting of the spectral densities between heavy-light parity partners must be driven by order parameters of spontaneous chiral symmetry breaking only.

Probing the chiral symmetry restoration via the change of order parameters requires a reliable extraction of their medium dependence.
As is long known for the vacuum case, the chiral condensate is numerically suppressed in QCD sum rules in the light quark sector due to the tiny light quark mass, but occurs amplified by the large heavy quark mass in QCD sum rules involving a light and a heavy quark \cite{Reinders:1984sr}.
Despite of the amplification in terms of $m_c \langle \bar u u \rangle$, with $m_c$ and $\langle \bar u u \rangle$ to be evolved to the appropriate scale, in case of the $D$ mesons, the dependence of the in-medium $D$ meson spectrum on the chiral condensate is not as direct as anticipated.
This is clear in so far as there are always particle and antiparticle contributions to the spectrum of pseudo-scalar states and one has to deal with both pieces.
This accounts for inherent suppressions and amplifications of different types of condensates due to the generic structure of the sum rule in that case.
Indeed, a precise analytic and numerical investigation points to competitive numerical impacts of various condensates for the mass splitting of particle and antiparticle.
Whereas, the determination of the mass center rather depends on the modeling of the continuum threshold \cite{Hilger:2008jg}.

In the difference of chiral partner spectra (Weinberg type sum rules) the dependence on chirally symmetric condensates drops out.
At the same time, the amplification of the chiral condensate by the heavy quark mass is still present.
This makes Weinberg type sum rules of mesons composed of heavy and light quarks an interesting object of investigation.

Our paper is organized as follows.
The operator product expansion (OPE) and a suitable projection are elaborated in section \ref{sct:delta_Pi}
for differences of current-current correlators.
These are spelled out for chiral partners of open charm mesons
in the form of Weinberg type sum rules with the focus on the OPE sides
in section \ref{sct:chiral_partners}; a remark on heavy-quark symmetry is included as well.
Numerical examples are exhibited in section \ref{sct:num}.
The summary can be found in section \ref{sct:summary}.
In Appendix \ref{sct:correlator}, 
some relations between current-current correlators are considered.
Appendices \ref{sct:app1} and \ref{sct:app_prop} list projection coefficients and prove a necessary relation for the quark propagator.


\section{Differences of current-current correlators and their operator product expansion}
\label{sct:delta_Pi}

We consider the currents
\begin{subequations}
\label{eq:def_current_2_flavor}
\begin{align}
j^{{\rm S}}(x) & :=  \bar q_1(x) \, q_2(x)  \: , \label{eq:def_scalar_2_flavor} \\
j^{{\rm P}}(x) & :=  \bar q_1(x) \, i\gamma_5 \, q_2(x) \: , \label{eq:def_pscalar_2_flavor} \\
j_\mu^{{\rm V}}(x) & :=  \bar q_1(x) \, \gamma_\mu \, q_2(x) \: , \label{eq:def_vector_2_flavor} \\
j_\mu^{{\rm A}}(x) & :=  \bar q_1(x) \, \gamma_5 \gamma_\mu \, q_2(x) \label{eq:def_avector_2_flavor}
\end{align}
\end{subequations}
and the corresponding causal correlators
\begin{subequations}\label{eq:def_corr_2_flavor}
\begin{align}
	\label{eq:corrdef_2_flavor_0}
	\Pi^{\rm (S,P)}(q) &= i \int \Dp{4}{x} e^{iqx} \langle {\rm T} \left[ j^{\rm (S,P)}(x) j^{\rm (S,P)\dagger}(0) \right] \rangle
	\: ,
	\\
	\label{eq:def_corr_2_flavor_1}
	\Pi^{\rm (V,A)}_{\mu\nu}(q) &= i \int \Dp{4}{x} e^{iqx} \langle {\rm T} \left[ j^{\rm (V,A)}_\mu(x) j^{\rm (V,A)\dagger}_\nu(0) \right] \rangle
	\: ,
\end{align}
\end{subequations}
where T$[\ldots]$ denotes time-ordering and $\langle \ldots \rangle$ means Gibbs averaging.
Retarded and advanced correlators are related via analytic continuation to the causal correlator \cite{Negele:1988vy}.

In the rest frame of the nuclear medium, i.e.\ $n=(1,\vec 0)$, and for mesons at rest, i.e.\ $q=(1,\vec 0)$, the (axial-) vector correlator can be decomposed as
\begin{equation}
\label{eq:proj_VA}
	\Pi^{\rm(V,A)}_{\mu\nu}(q) = \left( \frac{q_\mu q_\nu}{q^2} - g_{\mu\nu}\right) \Pi^{\rm(V,A)}_{\rm T}(q)
		+ \frac{q_\mu q_\nu}{q^2} \Pi^{\rm(V,A)}_{\rm L}(q)
\end{equation}
with
$
	\Pi^{\rm(V,A)}_{\rm T}(q) = \frac{1}{3} \left( \frac{q^\mu q^\nu}{q^2} - g^{\mu\nu} \right) \Pi^{\rm(V,A)}_{\mu\nu}(q)
$ and 
$
	\Pi^{\rm(V,A)}_{\rm L}(q) = \frac{1}{q^2} q^\mu q^\nu \Pi^{\rm(V,A)}_{\mu\nu}(q)
$ written explicitly covariant.
$\Pi_T$ contains the information about (axial-) vector degrees of freedom.
$\Pi_L$ refers to (pseudo-) scalar states and can be related to the (pseudo-) scalar correlator (see Appendix A for details).

\begin{figure}
\centerline{
\includegraphics[width=\textwidth]{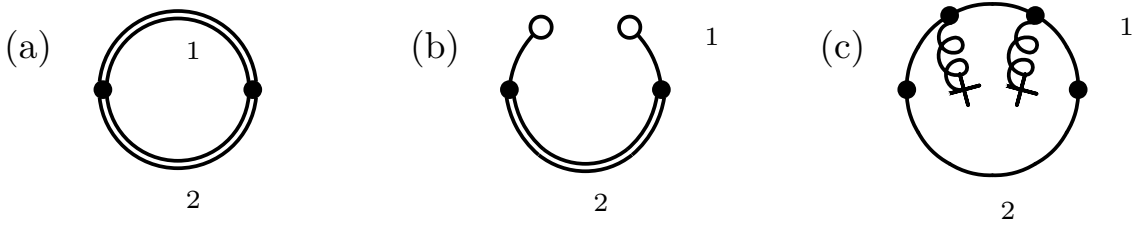}
}
\caption{Diagrammatical representation of (a) $\Pi^{(0)}$ 
and (b) $\Pi^{(2)}$. (c) is for $\Pi^{(0)}$ 
with two gluon lines attached to the free propagator of quark 1. 
Double lines stand for the complete perturbative series of the quark propagator, 
wiggly lines are for gluons, and circles denote non-local quark condensates, 
while crosses symbolize local quark or gluon condensates.}
\label{fig:cond}
\end{figure}

We now proceed with the OPE's for
$\Pi^X$ $\in$ $\left\{ \Pi^{\rm (S,P)}, \Pi^{\rm (V,A)} \equiv g^{\mu\nu} \Pi^{\rm (V,A)}_{\mu\nu} \right\}$.
We do not include the $\alpha_s$ corrections which would arise from inserting the next-to-leading order interaction term in the time-ordered product of the current-current correlator.
Such terms account, e.g., for four-quark condensates. They are of mass dimension 6 and beyond the scope of this investigation.
According to standard OPE techniques (see e.g.\ \cite{Novikov:1983gd,Shifman:1978by,Narison:2002pw}), 
the time-ordered product can be expanded into normal-ordered products 
multiplied by Wick-contracted quark-field operators.
Dirac indices can be projected onto elements of the Clifford algebra,
$\Gamma \in \{ \mathds{1}, \gamma_\mu, \sigma_{\mu\nu}, \gamma_5 \gamma_\mu, \gamma_5\}$, 
which provides an orthonormal basis
in the space of $4 \times 4$ matrices 
with the scalar product $(A,B) \equiv \frac14 \TrD[AB]$.
Color indices can be projected onto an analogously appropriate basis.
Thereby, color and Dirac traces of the quark propagator occur. 
Using the background field method in fixed-point gauge
(cf.~\cite{Novikov:1983gd,Shifman:1978by}) 
for the gluon background field $A_\mu(x)$, $x_\mu A^\mu(x) = 0$, 
one can expand the non-local quark field operators by a covariant expansion
$
q(x) = \sum_n \frac{x^{\alpha_1} \ldots x^{\alpha_n}}{n!}
        \left. \left( {D}_{\alpha_1} \ldots 
        {D}_{\alpha_n} q \right) \right|_{x=0}
$
to arrive, after a Fourier transformation, at
\begin{multline}
\label{eq:contr}
	\Pi^X (q)
	= -i^3\int \frac{\Dp{4}{p}}{(2\pi)^4} \langle : \TrCD \left[ S_1(p) \Gamma^X S_2(q+p) \Gamma^X \right] : \rangle
	\\
	+ i^2 \sum_\Gamma \frac 14 \sum_{n=0}^\infty \frac{1}{n!}
	\left( -i \right)^n \partial_q^{\vec\alpha_n}
	\langle : \bar{q}_1 \Gamma
	\TrD \left[ \Gamma \Gamma^X S_2(-q) \Gamma^X \right]
	{D}_{\vec{\alpha}_n} q_1
	\\
	\shoveright{
	+ (-1)^n \bar q_2 \Gamma \TrD \left[ \Gamma \Gamma^X S_1(q) \Gamma^X \right]
	{D}_{\vec{\alpha}_n} q_2 : \rangle
	}
	\\
	\shoveright{
    = \Pi^{X(0)}(q) + \Pi^{X(2)}(q)
    \: ,
    }
\end{multline}
where ${D}_{\vec{\alpha}_n} = D_{\alpha_1} \ldots D_{\alpha_n}$ (with an analog notation for the partial derivative) and quark fields and their derivatives are taken at $x=0$.
For $X$ denoting vector and axial-vector states,
$\left(\Gamma^X \right)_{ij}\left(\Gamma^X \right)_{kl} \equiv \left(\Gamma^X_\mu \right)_{ij}\left(\Gamma^{X\mu} \right)_{kl}$
is understood.
$\Pi^{(0)}(q)$ denotes the fully contracted (depicted in 
Fig.~\ref{fig:cond} (a)) and $\Pi^{(2)}(q)$ the 2-quark term (see Fig.~\ref{fig:cond} (b) 
for $\langle \bar \psi \ldots \psi \rangle$), $\TrCD$ means trace w.r.t.\ color and Dirac indices.
In doing so, we introduce the momentum-space perturbative quark propagator 
in a gluonic background field as
$ 
S(q) = \int \Dp{4}{x} \, e^{iqx} S(x,0) = \int \Dp{4}{x} \, e^{-iqx} S(0,x)
$
and $S(x,y) = -i \langle 0 | {\rm T} \left[ q(x) \bar q(y) \right] | 0 \rangle$.
To have the same structures in both quantities $\Pi^{(0)}$ and $\Pi^{(2)}$, 
we also project the matrix product $\Gamma_\mu^X S \Gamma^\mu_X$ 
onto this basis (see Appendix \ref{sct:app1}):
\begin{subequations} \label{eq:2}
\begin{align}
    & \Pi^{{\rm (P,S)}(0)}(q) = -i \int \frac{\Dp{4}{p}}{(2\pi)^4} \langle \ld \frac14 \TrC \left[
        \TrD[S_2(p+q)] \TrD[S_1(p)]
        \right.        \label{eq:2a}
        \\ & \quad
        + (-1)^{{\rm (P,S)}} \TrD[S_2(p+q) \gamma_\mu] \TrD[S_1(p)\gamma^\mu]
        \nonumber
        \\ & \quad
        + \frac12 \TrD[S_2(p+q)\sigma_{\mu\nu}] \TrD[S_1(p) \sigma^{\mu\nu}]
        \nonumber
        \\ & \quad
        + (-1)^{{\rm (P,S)}} \TrD[S_2(p+q) \gamma_5 \gamma_\mu] \TrD[S_1(p) \gamma_5 \gamma^\mu]
        \nonumber
        \\ & \quad
        \left.
        + \TrD[S_2(p+q) \gamma_5] \TrD[S_1(p)\gamma_5]
        \right] \rd \rangle \: , \nonumber
        \displaybreak[0]
    \\
    & \Pi^{{\rm (V,A)}(0)}(q) = i \int \frac{\Dp{4}{p}}{(2\pi)^4} \langle \ld \TrC \left[
        \TrD[S_2(p+q)] \TrD[S_1(p)]
        \right.        \label{eq:2b}
        \\ & \quad
        + (-1)^{{\rm (V,A)}} \TrD[S_2(p+q) \gamma_\mu] \TrD[S_1(p)\gamma^\mu]
        \nonumber
        \\ & \quad
        - (-1)^{{\rm (V,A)}} \TrD[S_2(p+q) \gamma_5 \gamma_\mu] \TrD[S_1(p) \gamma_5 \gamma^\mu]
        \nonumber
        \\ & \quad
        \left.
        - \TrD[S_2(p+q) \gamma_5] \TrD[S_1(p)\gamma_5]
        \right] \rd \rangle \: , \nonumber
\end{align}
\end{subequations}
where $(-1)^{\rm (P,V)} = -1$ for pseudo-scalar and vector mesons and $(-1)^{\rm (S,A)} = 1$ for scalar and axial-vector mesons.
$\Pi^{X(2)}(q)$ in \eqref{eq:contr} may be simplified using Tab.~\ref{tb:app2}
in Appendix \ref{sct:app1}.

To obtain the OPE's for the difference of chiral partners 
$\Pi^{\rm P-S} \equiv \Pi^P-\Pi^S$ and $\Pi^{\rm V-A} \equiv \Pi^V-\Pi^A$, 
one can use \eqref{eq:2} or directly project the occurring anti-commutators.
The result reads
\begin{subequations} \label{eq:3}
\begin{multline}
	\label{eq:3a}
    \Pi^{{\rm P-S}(0)}(q) = -i \int \frac{\Dp{4}{p}}{(2\pi)^4} \langle \ld \frac12 \TrC \left\{
        \TrD[S_2(p+q)] \TrD[S_1(p)]
        \right.
        \\
        + \frac12 \TrD[S_2(p+q)\sigma_{\mu\nu}] \TrD[S_1(p) \sigma^{\mu\nu}]
        \left.
        + \TrD[S_2(p+q) \gamma_5] \TrD[S_1(p)\gamma_5]
        \right\} \rd \rangle \: ,
\end{multline}
\begin{multline}
    \label{eq:3b}
    \Pi^{{\rm P-S}(2)}(q) = \sum_n \frac{(-i)^n}{n!} \frac12 \sum_\Gamma^{\{\mathds{1}, \sigma_{\alpha < \beta}, \gamma_5 \} }
        \langle \ld
        \overline{q}_1 \overleftarrow{D}_{\vec\alpha_n}
        \Gamma
        \partial^{\vec\alpha_n}
        \left( \TrD[ \Gamma S_2(q)] \right) q_1
        \\
        + \overline{q}_2 \Gamma \partial^{\vec\alpha_n}
        \left( \TrD[ \Gamma S_1(-q)] \right)
        \overrightarrow{D}_{\vec\alpha_n} q_2
        \rd \rangle
\end{multline}
\begin{multline}
    \label{eq:3c}
    \Pi^{{\rm V-A}(0)}(q) = i \int \frac{\Dp{4}{p}}{(2\pi)^4} \langle \ld 2 \TrC \left\{
        \TrD[S_2(p+q)] \TrD[S_1(p)]
        \right.
        \\
        \left.
        - \TrD[S_2(p+q) \gamma_5] \TrD[S_1(p)\gamma_5]
        \right\} \rd \rangle \: ,
\end{multline}
\begin{multline}
    \label{eq:3d}
    \Pi^{{\rm V-A}(2)}(q) = - \sum_n \frac{(-i)^n}{n!} 2 \sum_\Gamma^{\{\mathds{1},i\gamma_5\}}
        \times \langle \ld
        \overline{q}_1 \overleftarrow{D}_{\vec\alpha_n}
        \Gamma
        \partial^{\vec\alpha_n}
        \left( \TrD[ \Gamma S_2(q)] \right) q_1
        \\
        + \overline{q}_2 \Gamma \partial^{\vec\alpha_n}
        \left( \TrD[ \Gamma S_1(-q)] \right)
        \overrightarrow{D}_{\vec\alpha_n} q_2
        \rd \rangle \: ,
\end{multline}
\end{subequations}
where the Clifford basis is modified now by the imaginary unit in front 
of $\gamma_5$ for the vector--axial-vector difference.
The advantage is that we are left with three different types 
of Dirac traces for the quark propagators in the P$-$S case 
and only two in the V$-$A case.


The perturbative series for a momentum-space quark-propagator 
in a gluon background field in fixed-point gauge is presented 
in Appendix \ref{sct:app_prop}.
Its derivative is given by the Ward identity
$
\partial^\mu S(q) = - S(q) \Gamma^\mu(q,q;0) S(q) ,
$
where $\Gamma^\mu(q,q;0)$ denotes the exact quark-gluon vertex function 
at vanishing momentum transfer. $\Gamma^\mu(q,q;0) = \gamma^\mu$ 
holds for a classical background field 
meaning that the Ward identity for the complete perturbation series 
has the same form as for free quarks \cite{Landau_QED}.

For the limit of a massless quark flavor attributed to $q_1$, $m_1 \to 0$, 
one can show (see Appendix \ref{sct:app_prop}) that
$\TrD[\Gamma S_1(q)] = 0$
for $\Gamma \in \{\mathds{1}, \sigma_{\mu\nu}, \gamma_5\}$
and $q^2 \neq 0$.
In this limit only the diagram in Fig.~\ref{fig:cond} (b) 
gives a contribution to $\Pi^{(2)}$.
The corresponding diagram for $1 \longleftrightarrow 2$ vanishes, because the sum over Dirac matrices in \eqref{eq:3} covers such elements where the corresponding traces vanish.
Hence, for chiral partner sum rules the often used approximation of a static quark, which results in vanishing heavy-quark condensates, is not necessary as their Wilson coefficients vanish in the limit of the other quark being massless.
This means that, if one is seeking the occurrence of quark condensates 
in lowest (zeroth) order of the strong coupling $\alpha_s$, 
at least one quark must have a nonzero mass. 
(Higher order interaction term insertions cause the occurrence of 
further quark condensates proportional to powers of $\alpha_s$,
cf.~\cite{Kapusta:1993hq,Leupold:2006bp,Leupold:2006ih} for examples.)
Hence, the structure of the OPE side of the famous Weinberg sum rules 
for light quarks w.r.t.\ quark condensates is shown in all orders 
of the quark propagators and quark fields.

For the completely contracted term $\Pi^{(0)}$ the situation is somewhat more involved.
In case of two light quarks, i.e.\ $m_{1,2} \to 0$, 
the limiting procedure and the momentum integration commute and, hence, 
$\Pi^{(0)} = 0$ is obvious, because all non-vanishing terms drop out in the chiral difference.
If one of the quarks has a nonzero mass, 
the limiting procedure and the momentum integration do not commute
due to the occurrence of infrared divergences \cite{Chetyrkin:1982zq,Tkachov:1999nk,Grozin:1994hd,Jamin:1992se,HilgerDA},
e.g.\ for the term depicted in Fig.\ \ref{fig:cond} (c) which is proportional to the gluon condensate.
As the integration domains
in \eqref{eq:2} and (\ref{eq:3a}, \ref{eq:3c}) involve momenta $p = 0$,
${\rm Tr} [\Gamma S(p,m)]$ does not converge uniformly for
$m \to 0$. Hence, the integration and the limit $m \to 0$
cannot be interchanged. In fact, a direct calculation of the
perturbative contribution to chiral partner OPE's shows that the
infrared divergences are the only remaining terms.
All finite terms cancel out in the chiral difference.
These divergences 
have to be absorbed by introducing condensates which are not normal ordered.
We demonstrate here explicitly their cancellation.
Up to mass dimension 5, the only product of traces which contributes 
to $\Pi^{(0)}$ is $\TrD[S_1] \TrD[S_2]$.
This is true for arbitrary quark masses.
Indeed, up to order $\alpha_s^1$ one can show that the contribution
$-\int \Dp{4}{p} \TrD[S_2^{(0)}(p+q)] \langle \ld \TrD[S_1^{(2)}(p)] \rd \rangle$,
which is the remaining term of \eqref{eq:2} after the limit 
$m_1 \to 0$ has been taken,
is canceled by
$
\int \Dp{4}{p} \TrD[S_2^{(0)}(q)] \langle \TrD[S_1^{(2)}(p)] \rangle
$.
The latter quantity is introduced by the definition of non-normal ordered condensates
\begin{equation} \label{eq:4}
\begin{split}
\langle \Omega | \bar{q} \hat{O} \left[D_\mu\right] q | \Omega \rangle = &
\langle \Omega | \ld \bar{q} \hat{O} \left[D_\mu\right] q \rd | \Omega \rangle
\\ &
- i \int \frac{\Dp{4}{p}}{(2\pi)^4}
\langle \Omega | \TrCD \left[ \hat{O} 
\left[-ip_\mu - i \tilde{A}_\mu \right] S_q(p) \right] | \Omega \rangle \: ,
\end{split}
\end{equation}
where $\hat{O}\left[ D_\mu \right]$ is an arbitrary operator depending 
on $D_\mu$, and $\tilde{A}_\mu$ denotes the Fourier transform of $A_\mu$,
resulting in a factor $\int \langle \TrD[S_1^{(2)}(p)] \rangle$ 
to the Wilson coefficient $\TrD[S_2^{(0)}(q)]$ of the chiral condensate in \eqref{eq:contr}.
Hence, in case of heavy-light systems, first the integration has to be performed, 
then one has to introduce non-normal ordered condensates according to \eqref{eq:4} 
and afterward the limit $m_1 \to 0$ can be taken.
In case of equal masses, $m_1=m_2$, the divergences cancel each other with virtue to
\begin{equation}
\int \Dp{4}{p} \left(
\TrD[S_1^{(2)}(p+q)] \TrD[S_2^{(0)}(p)] + 
\TrD[S_1^{(0)}(p+q)] \TrD[S_2^{(2)}(p)] \right) = 0 \: .
\end{equation}

If two heavy (static) quarks are considered, 
only $\Pi^{(0)}$ gives a contribution to the chiral OPE, 
whereas for two massless quarks, both $\Pi^{(0)}$ and $\Pi^{(2)}$ vanish.

Putting everything together this means that for light quarks
($m_{1,2} \ll \Lambda_{QCD}$)
in the chiral-difference OPE 
the corresponding traces and, therefore, the corresponding Wilson coefficients vanish, 
while for heavy quarks ($m_{1,2} \gg \Lambda_{QCD}$) the condensates vanish.
To obtain quark condensates in order $\alpha_s^0$ the two flavors 
must be of different mass scales, i.e.\ $q_1 \in \{u, d (,s)\}$ 
is a light quark and $q_2 \in \{c, b, t\}$ is a heavy quark.
Hence, to seek for condensates which are connected 
to chiral symmetry breaking as possible order parameters in order $\alpha_s^0$, 
a natural choice is to consider chiral partner mesons composed of a light 
and a heavy quark.
Given the above mentioned experimentally envisaged research programs at FAIR 
\cite{CBM,PANDA,Friman:2011zz} we focus on open charm mesons.
The presented formulas may be directly transferred to open bottom mesons
by $m_c \to m_b$. 

\section{Chiral partners of open charm mesons}
\label{sct:chiral_partners}

We now consider a light ($q_1 \equiv q$) and a heavy ($q_2 \equiv q_c$) quark entering the currents in Eq.\ \eqref{eq:def_current_2_flavor}.

\subsection{The case of P$-$S}

For the P$-$S case we consider the pseudo-scalar $D(0^-)$ 
($D^\pm$, $D^0$ and $\bar D^0$) and its scalar partners $D^\ast_0(0^+)$.
Of course, all the results also account for other heavy--light (pseudo-) scalar mesons.
(For open charm mesons $D_s$ which contain a strange quark, however, 
the limit $m_s \to 0$ may not be a good approximation and terms $\propto m_s$ 
should be taken into account as well.)

Up to and including mass dimension 5, after absorbing the divergences 
in non-normal ordered condensates, the OPE gets the following compact form
\begin{equation}
\begin{split}
     \Pi^{\rm P-S}(q) &\equiv \Pi^{\rm P}(q) - \Pi^{\rm S}(q) = \Pi^{{\rm P}(2)}(q) - \Pi^{{\rm S}(2)}(q)
    \\
    & = \sum_n \frac{(-i)^n}{n!} \sum_\Gamma^{\{\mathds{1}, \sigma_{\alpha < \beta}\} }
        \langle
        \bar q \overleftarrow{D}_{\vec\alpha_n}
        \Gamma
        \partial^{\vec\alpha_n}
        \TrD[ \Gamma S_c(q)] q(0)
        \rangle
        \: ,
\end{split}
\label{eq:pre_spsr}
\end{equation}
where the sum over the elements of the Clifford algebra does not contain $\gamma_5$ 
up to this mass dimension anymore.
To evaluate the condensates in \eqref{eq:pre_spsr} in terms of expectation
values of scalar operators, Lorentz indices have to be projected onto $g_{\mu\nu}$ 
and $\epsilon_{\mu\nu\kappa\lambda}$
in vacuum, whereas the medium four-velocity $n_\mu$ provides an additional structure
at finite densities and/or temperatures \cite{Jin:1992id}.
Hence, new condensates must be introduced which vanish in the vacuum.
Thereby, temperature and density dependences stem from Gibbs averages 
of medium specific operators.
The evaluation in the nuclear matter rest frame $n^\mu = (1, \vec 0 \,)$
for mesons at rest yields
\begin{multline} \label{eq:pssr}
    \Pi^{\rm P-S}(q_0)
    = 2 \langle \bar q q\rangle \frac{m_c}{q_0^2-m_c^2}
    - \langle\bar q g \sigma \mathrsfs{G} q \rangle 
    \frac{m_c q_0^2}{(q_0^2-m_c^2)^3}
    \\
    + \left[ \langle\bar q g \sigma \mathrsfs{G} q \rangle 
    - 8 \langle\bar q D_0^2 q \rangle \right]
    \frac{m_c q_0^2}{(q_0^2-m_c^2)^3}
\end{multline}
where we separated a medium-specific term 
(last line, $\langle\bar q g \sigma \mathrsfs{G} q \rangle 
- 8 \langle\bar q D_0^2 q \rangle \equiv \langle \Delta \rangle $) 
vanishing in vacuum, $g^2 = 4\pi \alpha_s$, $\sigma_{\mu\nu} = i\left[\gamma_\mu,\gamma_\nu\right]/2$, $\mathrsfs{G}_{\mu\nu} = i\left[D_\mu,D_\nu\right]$ and $q$ denotes either $d$ or $u$ quark field operators.
The Eq.\ \eqref{eq:pssr} reduces to the vacuum result \cite{Narison:2003td,Hayashigaki:2004gq,Reinders:1984sr} at zero density and temperature.
The condensates 
$\langle \bar q q \rangle$, 
$\langle \bar q g \sigma \mathrsfs{G} q \rangle$ and 
$\langle\bar q D_0^2 q \rangle$ 
may have different medium dependences.

An odd part of the OPE, i.e.
$\Pi(q_0) = \Pi^{\rm even}(q_0) + q_0 \Pi^{\rm odd}(q_0)$
with
$\Pi^{\rm even}(q_0) = (\Pi(q_0) + \Pi(-q_0))/2$
and
$\Pi^{\rm odd}(q_0) = (\Pi(q_0) - \Pi(-q_0))/(2q_0)$,
where both functions are even w.r.t.\ $q_0$, does not appear up to this mass dimension.
Although there is no $\gamma_\mu$ projection of the condensates for difference 
OPE's of chiral partner, 
it may arise from an odd number of derivatives in \eqref{eq:pre_spsr}.

As usual within a QCD sum rule analysis \cite{Novikov:1983gd,Shifman:1978by,Narison:2002pw}, 
a Wick rotation to Euclidean momenta, $q_0 = iQ$ has to be performed.
The OPE in the deep space-like region based on asymptotically free quark degrees of freedom can be expected to converge.
The OPE of the current-current correlation function in turn 
is related to the hadronic spectral density by an in-medium dispersion relation.
To evaluate the correlator at arbitrary values of the energy $q_0$ off the real axis, 
$\text{Im} \, q_0 \neq 0$, e.g.\ $q_0^2 = - Q^2 \to - \infty$, 
an integral over its discontinuities (spectral densities) along the real axis 
$\Delta \Pi(q_0, \vec{q} = 0) = \frac{1}{2i} \lim_{\epsilon \to 0} 
\left[ \Pi(q_0 + i\epsilon) - \Pi(q_0-i\epsilon) \right] = \text{Im} \Pi(q_0)$ 
is performed.
After a Borel transformation the result is
\begin{multline} \label{eq:psbsr}
    \frac{1}{\pi} \int_{-\infty}^{+\infty} \D{\omega} \, e^{-\omega^2/M^2} \omega \, \Delta \Pi^{\rm P-S}(\omega)
    =
    e^{-m_c^2/M^2}
    \Biggl[ - 2 m_c \langle \overline{q} q \rangle
    \Biggr.
    \\
    \Biggl.
    + \left( \frac{m_c^3}{2M^4} - \frac{m_c}{M^2} \right) \langle\overline{q} g \sigma \mathrsfs{G} q \rangle
    - \langle \Delta \rangle 
    \left( \frac{m_c^3}{2M^4} - \frac{m_c}{M^2} \right) \Biggr] \: .
\end{multline}
$M$ is the Borel mass. In a medium,
the integral over the spectral densities runs over positive and negative frequencies 
$\omega$, covering both particles and their antiparticles, of given quantum numbers.

It is instructive to cast Eq.~\eqref{eq:psbsr} in the form of Weinberg type sum rules 
\cite{Weinberg:1967kj,Kapusta:1993hq}.
This can be accomplished by expanding the exponential on both sides 
and comparing the coefficients of inverse powers of the Borel mass.
In such a way we can relate moments of the spectral P$-$S 
differences to condensates via
\begin{subequations} \label{eq:spmom}
\begin{align}
    \frac{1}{\pi} \int_{-\infty}^{+\infty} \D{\omega} \, \omega \Delta \Pi^{\rm P-S}(\omega) = & 
    - 2 m_c \langle \bar q q \rangle \: ,
    \\
    \frac{1}{\pi} \int_{-\infty}^{+\infty} \D{\omega} \, \omega^3 \Delta \Pi^{\rm P-S}(\omega) = &
    - 2 m_c^3 \langle \overline{q}q \rangle 
    + m_c \langle\bar q g \sigma \mathrsfs{G} q \rangle
    - m_c \, \langle \Delta \rangle  
    \\
    \frac{1}{\pi} \int_{-\infty}^{+\infty} \D{\omega} \, \omega^5 \Delta \Pi^{\rm P-S}(\omega) = &
    - 2 m_c^5 \langle \overline{q}q \rangle 
    + 3 m_c^3 \langle\overline{q} g \sigma \mathrsfs{G} q \rangle
    \label{eq:spmomc} 
    - 3 m_c^3 \, \langle \Delta \rangle 
    + \ldots, 
\end{align}
\end{subequations}
where the dots denote the neglected contribution of mass dimension 7.
This generalizes the OPE side of Weinberg type sum rules to scalar and pseudo-scalar mesons in the heavy-light quark sector for the first time.

If one attributes chiral symmetry to the degeneracy of chiral partners
(i.e.\ the l.h.s.\ of \eqref{eq:spmom} vanishes) 
the vanishing of $\langle \bar q q \rangle$ and
$\langle \bar q D_0^2 q \rangle = \frac 18 \left( \langle \bar q g \sigma \mathrsfs{G} q \rangle - \langle \Delta \rangle \right)$
on the r.h.s.\ is required. 
In this spirit, these condensates may be considered 
as possible order parameters of chiral symmetry.
Note that \eqref{eq:spmom} also allows to consider the omitted mass dimension 7 condensate as an order parameter.
Remarkably, the chiral condensate $\langle \bar q q\rangle$ of light quarks 
figures here in conjunction with the heavy quark mass as parameter 
for chiral symmetry breaking in each of the moments
(for vacuum, cf.~\cite{Narison:2003td}).
Of next importance is 
$\langle\bar{q} D_0^2 q \rangle$, 
again in combination with the heavy quark mass.
The r.h.s.\ quantities must be taken at a proper renormalization scale. 
Formally, in the chiral (i.e. strictly massless) limit for {\em all} quarks the 
r.h.s.\ of \eqref{eq:spmom} would vanish.


\subsection{The case V$-$A}

In the same manner we proceed in the V$-$A case.
From Eqs.~\eqref{eq:3c} and \eqref{eq:3d} 
we obtain up to and including mass dimension 5
\begin{equation} \label{eq:pre_vasr}
\begin{split}
    \Pi^{\rm V-A}(q) & \equiv \Pi^{\rm V}(q) - \Pi^{\rm A}(q) = \Pi^{{\rm V}(2)}(q) - \Pi^{{\rm A}(2)}(q)
    \\
   & = - \sum_n \frac{(-i)^n}{n!} 2 \sum_\Gamma^{\{\mathds{1},i\gamma_5\}}
        \langle
        \overline{q}_1 \overleftarrow{D}_{\vec\alpha_n}
        \Gamma
        \partial^{\vec\alpha_n}
        \TrD[ \Gamma S_2(q)] q_1
        \rangle
        \: ,
\end{split}
\end{equation}
where only the $\mathds{1}$-projection survives.
The in-medium evaluation results in 
\begin{equation} \label{eq:vasr}
\begin{split}
    \Pi^{\rm V-A}(q)
    = & - 8 \langle \bar q q \rangle \frac{m_c}{q_0^2-m_c^2}
    + 4 \langle\overline{q} g \sigma \mathrsfs{G} q \rangle \frac{m_c^3}{(q_0^2-m_c^2)^3}
    - 4 \langle \Delta \rangle 
    \frac{m_c q_0^2}{(q_0^2-m_c^2)^3}
\end{split}
\end{equation}
and together with \eqref{eq:vaps} and \eqref{eq:pssr} we obtain for the correlator containing the information about the vector and axial-vector degrees of freedom
\begin{equation}
\label{eq:va_ps_corr}
    \Pi^{\rm V-A}_{\rm T} (q)
    =
    \Pi^{\rm P-S}
    + \frac{m_c}{(q_0^2-m_c^2)^2} \langle \bar q g \sigma {\cal G} q \rangle 
    + \frac 13 \frac{m_c}{(q_0^2-m_c^2)^2} \langle \Delta \rangle
    \: .
\end{equation}
For vacuum, the result of \cite{Hayashigaki:2004gq,Reinders:1984sr} is recovered.
The Borel transformed sum rule is given by
\begin{multline} \label{eq:vabsr}
    \frac{1}{\pi} \int_{-\infty}^{+\infty} \D{\omega} \, e^{-\omega^2/M^2} \omega \, \Delta \Pi^{\rm V-A}_{\rm T}(\omega)
    = 
    e^{-m_c^2/M^2}
    \Biggl[ - 2 m_c \langle \overline{q} q \rangle
    \Biggr.
    \\
    \Biggl.
    + \frac{m_c^3}{2M^4} \langle \overline{q} g \sigma \mathrsfs{G} q \rangle
    - \langle \Delta \rangle 
    \left( \frac{m_c^3}{2M^4} - \frac 43 \frac{m_c}{M^2} \right) \Biggr] .
\end{multline}
The corresponding moments therefore are
\begin{subequations} \label{eq:vamom}
\begin{align}
    \frac{1}{\pi} \int_{-\infty}^{+\infty} \D{\omega} \, \omega \,\Delta \Pi^{\rm V-A}_{\rm T}(\omega) = & 
    -2 m_c \langle \bar q q \rangle \: ,
    \\
    \frac{1}{\pi} \int_{-\infty}^{+\infty} \D{\omega} \, \omega^3 \Delta \Pi^{\rm V-A}_{\rm T}(\omega) = &
    - 2 m_c^3 \langle \overline{q}q \rangle
    - \frac 43 m_c \langle \Delta \rangle
    \: ,
    \\
    \frac{1}{\pi} \int_{-\infty}^{+\infty} \D{\omega} \, \omega^5 \Delta \Pi^{\rm V-A}_{\rm T}(\omega) = &
    -2 m_c^5 \langle \bar q q \rangle
    + m_c^3 \langle \overline{q} g \sigma \mathrsfs{G} q \rangle
    -\frac{11}{3} m_c^3 \langle \Delta \rangle
    + \ldots
\end{align}
\end{subequations}
with the same meaning of ''$\ldots$'' as above.
This second set of Weinberg type sum rules contains the same condensates as the
first set in Eq.~(\ref{eq:spmom}) but in different combinations.
Again, addressing chiral symmetry to the l.h.s.\ one may consider the chiral condensate $\langle \bar q q \rangle$ and $\langle \Delta \rangle$ as possible order parameters.
As the P$-$S case allowed us to identify $\langle \bar q D_0^2 q \rangle$ as order parameter, $\langle \bar q g \sigma \mathrsfs{G} q \rangle$ also qualifies as order parameter.

In addition,
\begin{equation}
\label{eq:1stweinberg}
	\frac{1}{\pi} \int_{-\infty}^{+\infty} \D{\omega} \omega \Delta \tilde{\Pi}_{\rm T}^{\rm V-A}(\omega)
	= 0
\end{equation}
follows from \eqref{eq:proj_VA} but for the decomposition
$\Pi_{\mu\nu} = (q_\mu q_\nu - q^2 g_{\mu\nu}) \tilde{\Pi}_{\rm T} + q_\mu q_\nu \tilde{\Pi}_{\rm L}$.
This corresponds to Weinberg's first sum rule \cite{Weinberg:1967kj}.
Note that, in contrast to Weinberg's original sum rule \cite{Weinberg:1967kj}, no Goldstone boson properties appear on the right hand side of \eqref{eq:1stweinberg} because the heavy-light currents involved in our case are generally not conserved.
The Borel transformed sum rule for $\tilde{\Pi}_{\rm T} = \Pi_{\rm T}/q^2$ reads
\begin{multline}
\label{eq:vatldbsr}
	\frac{1}{\pi} \int_{-\infty}^{+\infty} \D{\omega} e^{-\omega^2/M^2} \omega \Delta \tilde{\Pi}_{\rm T}^{\rm V-A}(\omega)
	\\
	= \langle \bar q q \rangle \frac{2}{m_c} \left[1-e^{-m_c^2/M^2}\right]
	- \langle \overline{q} g \sigma \mathrsfs{G} q \rangle \frac{1}{m_c^3} \left[
	1-e^{-m_c^2/M^2}\left(1 + \frac{m_c^2}{M^2} + \frac{m_c^4}{2M^4}\right) \right]
	\\
	+ \langle \Delta \rangle \frac{1}{m_c^3} \left[
	\frac 73 \left[1-e^{-m_c^2/M^2}\left(1 + \frac{m_c^2}{M^2}\right)\right] -e^{-m_c^2/M^2} \frac{m_c^4}{2M^4} \right]
\end{multline}
and reproduces \eqref{eq:vamom}.
Note that using $\Pi_{\rm T}$, instead of $\tilde{\Pi}_{\rm T}$, is more appropriate for the heavy-quark limit.

\subsection{Heavy-quark symmetry}

In the heavy-quark limit, $m_2^2 \gg \left|q^2\right|$, the leading contributions in \eqref{eq:va_ps_corr} are
\begin{equation}
    \Bigl. \Pi^{\rm V-A}_{\rm T} (q) \Bigr|_{m_2^2 \gg \left| q^2 \right|}
    \approx
    \Bigl. \Pi^{\rm P-S}(q) \Bigr|_{m_2^2 \gg \left| q^2 \right|}
    \approx - \frac{2}{m_2} \langle \bar q q \rangle
    \: ,
\end{equation}
where
$\frac{1}{q^2-m_2^2} = -\frac{1}{m_2^2} \sum_{n=0}^\infty \left( \frac{q^2}{m_2^2} \right)^n$
has been exploited.
This result is in agreement with the expected degeneracy of vector and pseudo-scalar mesons and of axial-vector and scalar mesons in the heavy-quark limit \cite{Isgur:1989vq}.
Actually the $\langle \bar q q \rangle$ parts of $\Pi^{\rm P-S}$ and
$\Pi^{\rm V-A}_{\rm T}$ agree as one can check by comparing \eqref{eq:psbsr} with \eqref{eq:vabsr}.

\subsection{Numerical examples}
\label{sct:num}

\begin{figure}
\begin{center}
\includegraphics[width=.5\textwidth]{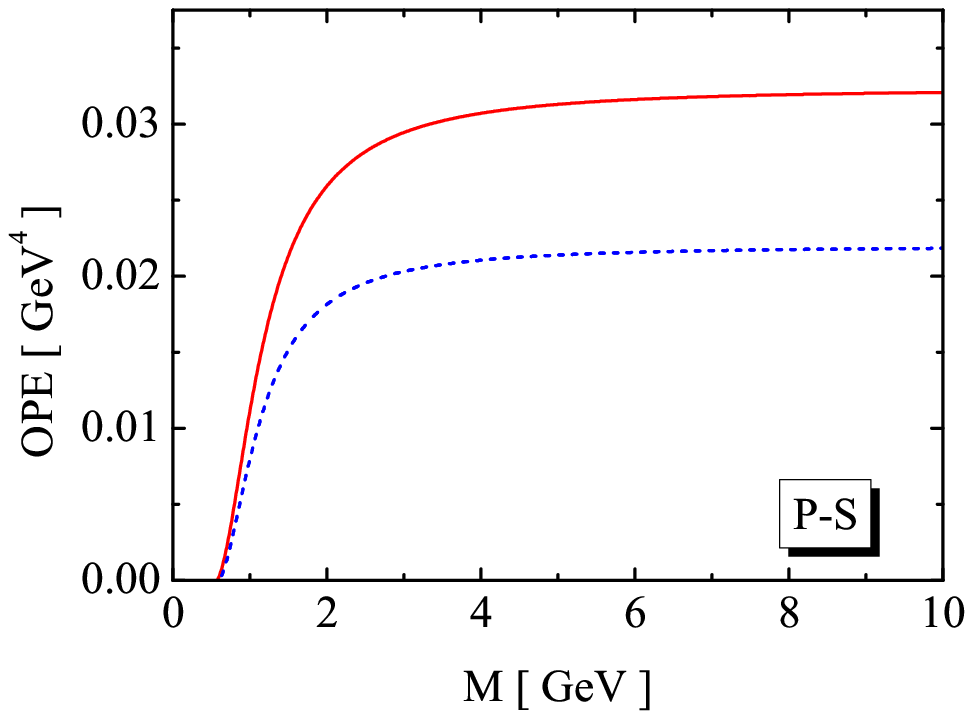}
\includegraphics[width=.5\textwidth]{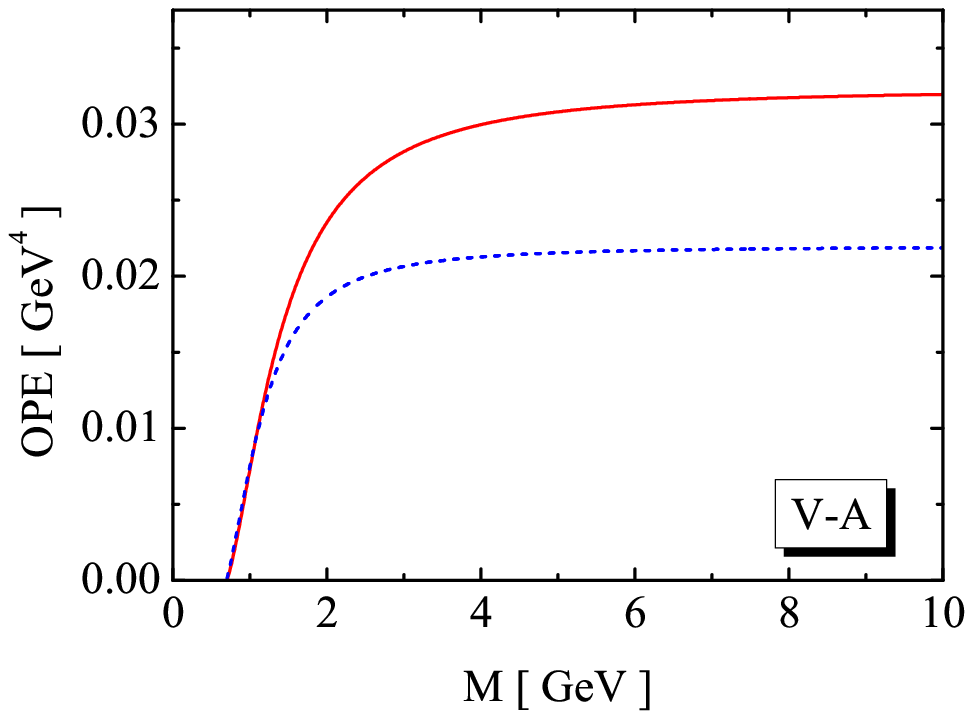}
\includegraphics[width=.5\textwidth]{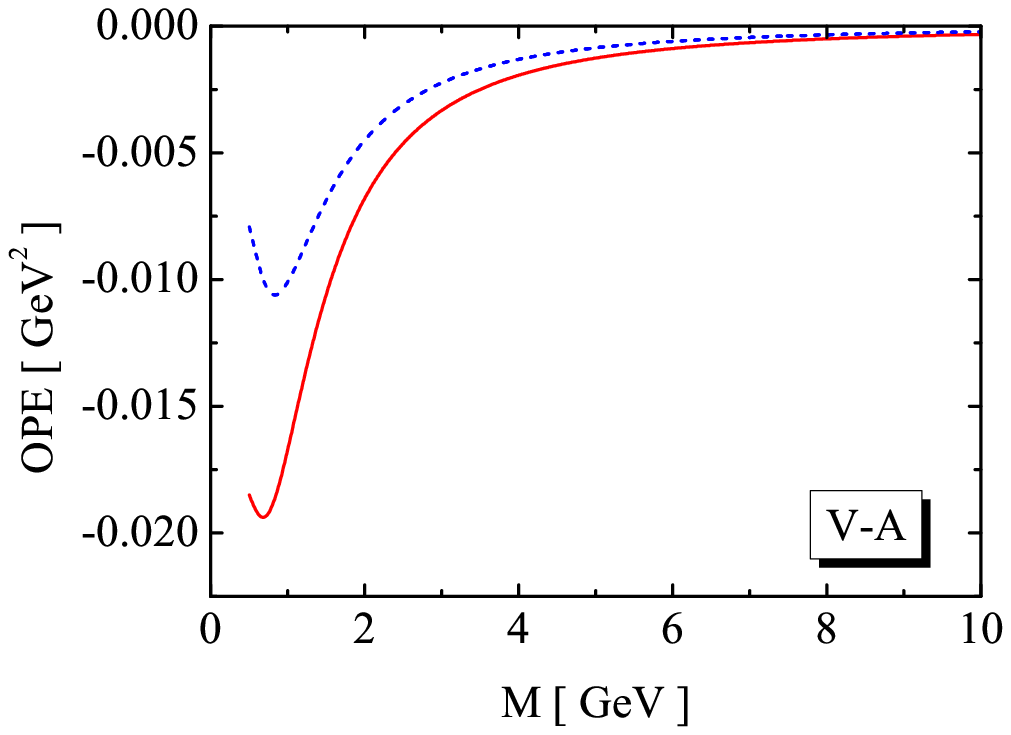}
\caption{
The OPE sides of Eqs.~\eqref{eq:pssr} (upper panel, P$-$S), \eqref{eq:vabsr}
(middle panel, V$-$A) and \eqref{eq:vatldbsr} (lower panel V$-$A)
as a function of the Borel mass in vacuum (solid red line) and at nuclear saturation density $n=0.15$ fm$^{-3}$ (dashed blue line).
The employed condensates are 
$\langle \bar q q \rangle = \langle \bar q q \rangle_0 + 45 n/11$,
$\langle \bar q  g \sigma \mathrsfs{G} q \rangle = 0.8 \, {\rm GeV}^2 
\langle \bar q q \rangle_0 + 3 n$ GeV$^2$, 
$\langle \bar q q \rangle_0 = (-.245 \,{\rm GeV})^3$ 
and 
$\langle \Delta \rangle = 8 \times 0.3 \, n$ GeV$^2$.
}
\label{fig:rhs}
\end{center}
\end{figure}

Using the condensates from \cite{Hilger:2008jg}
the r.h.s\ of Eqs.~\eqref{eq:pssr}, \eqref{eq:vabsr} and \eqref{eq:vatldbsr},
i.e.\ the OPE sides, are exhibited in Fig.~\ref{fig:rhs} 
as a function of the Borel mass for vacuum and
cold nuclear matter.
The Borel curves are significantly modified by changes of the 
entering condensates due to their density dependences, i.e.\ their magnitudes are lowered by approximately 30\% of their vacuum values.
It should be 
emphasized, however, that the density dependence is estimated
in linear approximation.
For the chiral condensate 
$\langle \bar q q \rangle$ it is known \cite{Lutz:1999vc,Kaiser:2007nv} that the actual
density dependence at zero temperature is not so strong. One may expect accordingly
a somewhat weaker impact of the medium effects.

For further evaluations, the hadronic spectral functions or moments
thereof must be specified, e.g.\ by using suitable moments
as in \cite{Zschocke:2002mn,Thomas:2005dc}.
 
\section{Summary}
\label{sct:summary}

In summary, we present difference QCD sum rules for chiral partners of mesons
with a simple quark structure.
Focussing on the operator product expansion of the current current correlator 
in lowest (zeroth) order of an interaction term insertion,
only the combination of a light and a heavy (massive) quark yields a non-trivial
result: Differences of spectral moments between pseudo-scalar and scalar
as well as vector and axial-vector mesons for condensates up to mass dimension $\leq 5$ are determined by
the combinations
$m_c \langle \bar q q \rangle$,
$m_c \langle \bar q  g \sigma \mathrsfs{G} q \rangle$, and
$m_c (\langle\bar q g \sigma \mathrsfs{G} q \rangle - 8 \langle\bar q D_0^2 q \rangle)$
(to be taken at an appropriate scale)
which may be considered as elements of ''order parameters'' 
of chiral symmetry breaking (see \cite{Doi:2004jp} for a lattice evaluation of the mixed quark-gluon condensate at finite temperature).
Vanishing of these condensates at high baryon density and/or temperature
would mean chiral restoration, i.e.\ the degeneracy of spectral moments of the
considered chiral partners.  
Chiral partners of mesons with light--light or heavy--heavy quark currents
are non-degenerate in higher orders of $\alpha_s$,
as exemplified by the Kapusta-Shuryak sum rule. 
The famous Weinberg sum rules generalized to a hot medium in \cite{Kapusta:1993hq}
refer to lower mass dimension moments where the OPE side vanishes in the
chiral limit.

Our results show a significant change of the OPE side when changing from vacuum to normal nuclear matter density.
This implies that also the hadronic spectral functions may experience a significant reshaping in the medium.


The open charm meson sector will be investigated in precision experiments
by two collaborations
in near future at FAIR in proton and anti-proton reactions at nuclei
as well as in heavy-ion collisions.
These experiments will shed some light on the mass spectrum of open charm mesons and its medium modifications.

{\it Acknowledgments:}
The work is supported by GSI-FE and BMBF 06DR9059.


\begin{appendix}

\section{Relations between current-current correlators}
\label{sct:correlator}

We denote the currents in Eq.\ \eqref{eq:def_current_2_flavor} by
\begin{equation} \label{eq:def_gen_curr}
j^{X,\tau} = \bar\psi \, \Gamma^X \tau \, \psi \: , \qquad \Gamma^X \in \left\{ \mathds{1}, i \gamma_5, \gamma_\mu, \gamma_5 \gamma_\mu \right\}\,.
\end{equation}
where $\psi =(q_1,\ldots,q_{n_f})$ collects the fields of $n_f$ quark flavors and $\tau$ is
an arbitrary matrix in flavor space.
Due to the special choice of $\Gamma^X$ the current fulfills
$\left( j^{X,\tau}(x) \right)^\dagger 
	= j^{X,\tau^\dagger}(x) \equiv \bar \psi \Gamma^X \tau^\dagger \psi
$.
Lorentz indices are suppressed here for V and A currents.

The currents $j^S$ and $j^P$ are supposed to carry the quantum numbers of scalar and pseudo-scalar mesons, respectively. The issue
is more complicated for $j^V$ and $j^A$.
To see this we consider the causal correlator of two vector-field valued, mesonic (bosonic) Heisenberg operators $A_\mu$ and $B_\nu$
(to be identified with $j^{\rm (V,A),\tau}_\mu(x)$ and $j^{\rm (V,A),\tau^{\dagger}}_\nu(y)$)
\begin{equation}
\label{eq:corrdef}
	\Pi_{\mu\nu}(q) = i \int \Dp{4}{x} e^{iqx} \langle {\rm T} \left[ A_\mu(x) B_\nu(0) \right] \rangle
	\: .
\end{equation}

A system at nonzero temperature and/or (baryon) density is characterized by a heat-bath vector $n_\mu$ (encoded in the Gibb's averages in \eqref{eq:def_corr_2_flavor})
which can be normalized by $n^2=1$.
The heat bath is at rest in a reference frame with $\vec n =0$.
For a system with a conserved current $j^B_\mu$, e.g., the baryon current, one can use $n_\mu \propto j^B_\mu$ (Eckart choice).
For a thermal system without conserved currents one must lock
the flow with local energy density (Landau choice, cf.\ \cite{Rangamani}).

A rank-2 tensor in four dimensional Minkowski space-time can be decomposed into six independent algebraic invariants $A, \ldots, F$ being functions of $q^2$, $n\cdot q$:
\begin{multline}
\Pi_{\mu\nu}(q) =
	\left( g_{\mu\nu} - \frac{q_\mu q_\nu}{q^2} \right) \, A + q_\mu q_\nu B
	+ n^q_\mu n^q_\nu  C
	\\
	+ (q_\mu n^q_\nu + n^q_\mu q_\nu) \, D
	+ (q_\mu n^q_\nu - n^q_\mu q_\nu) \, E
	+ \epsilon_{\mu\nu}^{\phantom{\mu\nu}\alpha\beta} q_\alpha n^q_\beta \, F
\end{multline}
where
$
	n_\mu^q := n_\mu - \frac{n \cdot q}{q^2} \, q_\mu
$.
From invariance w.r.t.\ parity one obtains $F=0$.
Imposing aAdditionally, time reversal invariance together with translational invariance of the considered medium, e.g.\ nuclear matter, give rise to the symmetry property $\Pi_{\mu\nu} = \Pi_{\nu\mu}$, hence, $E=F=0$.
If the correlator (causal, advanced, retarded correlator or spectral density) was four-transversal in both indices, i.e.\ $q^\mu \Pi_{\mu\nu} = q^\nu \Pi_{\mu\nu} = 0$, $E=0$ would follow without additional symmetry constraints.
Current conservation alone is not a sufficient requirement for the transversality of the causal correlator.

The coefficients $A$ and $C$ carry the information about the four-transversal degrees of freedom
referring to vector or axial-vector components
which can be decomposed into three-momentum transversal and three-momentum longitudinal states \cite{Gale:1990pn}.
The coefficient $B$ is related to the four-longitudinal states
referring to a scalar or pseudo-scalar component.
The coefficient $D$ encodes the mixing between three-momentum longitudinal (axial-) vector states and (pseudo-) scalar states
occurring due to broken rotational invariance for an excitation (hadron) moving with nonzero velocity in the medium \cite{Wolf:1997iba,Chin:1977iz}.
Thus, the spin is no longer a conserved quantum number.

In the special case $n = (1,\vec 0)$ and $q=(q_0, \vec{0})$, one has $n_\mu^q = 0$ and, therefore, $C$ and $D$ drop out.
Hence, the correlator can be decomposed into a longitudinal ($\Pi_{\rm T} = -A$) and a transverse ($\Pi_{\rm L} = q^2 B$) part \cite{Reinders:1984sr}, see Eq.\ \eqref{eq:proj_VA}.
At zero density and zero temperature $ \Pi_{\rm L} $ would vanish for $B_\nu$ being conserved as demonstrated below.

To relate the four-longitudinal part of the (axial-) vector current to the (pseudo-) scalar current let us consider the correlator
\begin{equation}
	\Pi_{\mu\nu}(q,p) = i^2\int \Dp{4}{x} \Dp{4}{y} e^{iqx} e^{-ipy} \langle {\rm T} \left[ A_\mu(x) B_\nu(y) \right] \rangle
	\: .
\end{equation}
Translation invariance allows to factor out a momentum dependence
\begin{equation}
\label{eq:corr1}
\begin{split}
	\Pi_{\mu\nu}(q,p)
		& = i (2\pi)^4 \delta^{(4)}(p-q) \Pi_{\mu\nu}(q)
		\: .
\end{split}
\end{equation}
Contracting with $q_\mu p_\nu$, performing a suitable number of partial integrations in each term and using
$\partial_x^0 A_0(x) = \partial_x^\mu A_\mu(x) - \partial_x^i A_i(x)$
gives the non-anomalous Ward identity relating the longitudinal part of the causal correlator of two Lorentz vectors to the correlator of their divergences
\begin{multline}
\label{eq:corr2}
q^\mu p^\nu \Pi_{\mu\nu} (q,p)
        = - i^2 \int \Dp{4}{x} \Dp{4}{y} e^{iqx} e^{-ipy} \langle
        {\rm T} \left[ i \partial_x^\mu A_\mu(x) i \partial_y^\nu B_\nu(y) \right]
        \\
        +i \delta(x_0-y_0) \left( q^\mu \left[A_\mu(x), B_0(y) \right]_{x_0=y_0}
        + \left[ A_0(x), i \partial_y^\nu B_\nu(y) \right]_{x_0=y_0}
        \right)
        \rangle
        \: .
\end{multline}
The first and third terms on the right hand side of this equation vanish if $\partial^\nu B_\nu = 0$ holds.

Identifying $A_\mu$ and $B_\nu$ with the currents above,
$j_\mu^{{\rm V},\tau}(x) := \bar\psi(x) \, \gamma_\mu \, \tau \, \psi(x)$, 
$j_\mu^{{\rm A},\tau}(x) := \bar\psi(x) \, \gamma_5 \gamma_\mu \, \tau \, \psi(x)$, 
and using the equations of motion for the quark fields, their divergences are given by
\begin{equation}
	i \partial^\mu j_\mu^{{\rm (V,A)}, \tau} = \bar \psi \left( \gamma_5 \right) \left[ \tau, M \right]_{\mp} \psi
	= (-i)^{\rm (S,P)} j^{{\rm (S,P)}, \left[\tau,M\right]_\mp}
	\: ,
\end{equation}
where the upper (lower) sign corresponds to the vector (axial-vector) current and
$M = {\rm diag} \left( m_1, \ldots, m_{n_f} \right)$ is the quark mass matrix.

The equal-time commutators in \eqref{eq:corr2} can be obtained from the field anti-commutators or, equivalently, from current algebra considerations for the first commutator as the second involves a divergence.
The equal-time current commutation relations
are given by
\begin{subequations} \label{eq:etc}
\begin{align}
	\left[ j_\mu^{{\rm (V,A)},\tau} (x), \left( j_0^{{\rm (V,A)},\tau^\prime} (y) \right)^\dagger \right]_{x_0 = y_0}
		& = \delta^{(3)}(\vec x - \vec y) j^{{\rm V}, \left[ \tau, \tau^{\prime\dagger} \right]}_\mu(x)
		\: ,
		\\
	\left[ j_0^{{\rm (V,A)},\tau} (x), i \partial^\nu \left( j_\nu^{{\rm (V,A)},\tau^\prime} (y) \right)^\dagger \right]_{x_0 = y_0}
		& = \mp \delta^{(3)}(\vec x - \vec y) j^{{\rm S}, \left[ \tau, \left[ M,\tau^{\prime\dagger}\right]_\mp\right]_\mp}(x)
		\: .
\end{align}
\end{subequations}
We here consider commutators of currents with the same parity.
Setting $A_\mu(x) = j^{{\rm (V,A)},\tau}_\mu(x)$ and $B_\nu(y) = j^{{\rm (V,A)},\tau^{\prime\dagger}}_\nu(y)$ in \eqref{eq:corr2}, inserting the equal time commutators \eqref{eq:etc}, using translational invariance and comparing with \eqref{eq:corr1} we obtain the non-anomalous Ward identity which relates the causal correlator of (axial-) vector currents to the causal correlator of their divergences
\begin{multline}
\label{eq:gen_vasp}
	q^\mu q^\nu \Pi_{\mu\nu}^{\rm (V,A)}\left(q \left|\tau,\tau^\prime \right. \right)
	\\
	\shoveright{
	= \Pi^{\rm (S,P)} \left( q \left| \left[\tau,M\right]_\mp, \left[\tau^\prime,M\right]^\dagger_\mp \right. \right)
	- \langle q^\mu j_\mu^{{\rm V},\left[\tau,\tau^{\prime\dagger}\right]} \rangle
	\mp \langle j^{{\rm S},\left[\tau,\left[M,\tau^{\prime\dagger}\right]_\mp \right]_\mp}
	\rangle
	\: .
	}
\end{multline}


We turn now to an $n_f = 2$ flavor system.
Let heavy-light meson currents be given by
$\tau = \tau^\prime = (\sigma^1 + i \sigma^2)/2$, where $\sigma^i = 2 t^i$ are the Pauli matrices,
which gives \eqref{eq:def_current_2_flavor}.
The diagonal mass matrix can be written as
$M = (C_{\rm V} \sigma_3 + C_{\rm A} \mathds{1})/2$
and the commutators fulfill
$[\tau,M]_\mp = \mp C_{\rm (V,A)} \tau$,
$\left[\tau,\tau^\dagger\right]_\mp = \tau^{\rm (V,A)}$,
$\left[\tau, \left[M,\tau^\dagger\right]_\mp \right]_\mp = \mp C_{\rm (V,A)} \tau^{\rm (V,A)}$,
where we have defined
$C_{\rm V} = m_1 - m_2$, $C_{\rm A} = m_1 + m_2$
and $\tau^{\rm V} = \sigma^3$, $\tau^{\rm A} = \mathds{1}$.
Note that $q_1$ is attributed to a light-quark field (e.g. up or down) and $q_2$ to a heavy-quark field (e.g. charm or bottom).

With these relations the longitudinal part of \eqref{eq:proj_VA} is given as
\begin{equation}
\label{eq:corrvasp}
    q^2 \Pi^{\rm (V,A)}_{\rm L}(q) = q^\mu q^\nu \Pi^{\rm (V,A)}_{\mu\nu}(q) = 
    C_{\rm (V,A)}^2 \Pi^{\rm (S,P)}(q) - \langle \bar \psi \hat q \sigma^3 \psi \rangle
        + C_{\rm (V,A)}
        \langle \bar \psi \tau^{\rm (V,A)} \psi \rangle
\end{equation}
with $\hat q = \gamma_\mu q^\mu$.
(Equation \eqref{eq:corrvasp} also holds in a three-flavor system with two light quarks and one massive quark.)
Furthermore $\Pi_{\rm L}$ is zero if the current $j^{X^\prime,\tau^\prime}_\nu(y)$ (or $B_\nu$ in \eqref{eq:corrdef}) is conserved, i.e.\ $m_1=m_2$ for vector currents or $m_1+m_2=0$ for axial-vector currents (first and third term on the right hand side of Eq.\ \eqref{eq:corrvasp} vanish), and the difference of light and heavy net quark currents being zero, i.e.\
$\langle \bar \psi \hat q \sigma^3 \psi \rangle = 0$.
Current conservation alone is not sufficient for a vanishing longitudinal projection.
Indeed, the transversality in case of current conservation generally reads
$
	q^\mu \Pi_{\mu\nu}^{\rm (V,A)}\left( q \left| \tau, \tau^\prime \right. \right)
		= - \langle j_\nu^{\rm V} \left( 0 \left| \left[\tau, \tau^{\prime\dagger} \right]_- \right. \right) \rangle
$,
which only vanishes in a medium which is symmetric w.r.t.\ the quark flavors of the meson current.
If $\Pi_{\rm L}$ is zero, the transversal projection $\Pi^{\rm (V,A)}_{\rm T}(q)$ is proportional to the trace of the correlator
$\Pi^{\rm (V,A)}_{\rm T}(q) = - g^{\mu\nu} \Pi^{\rm (V,A)}_{\mu\nu}(q) / 3$.
Otherwise the trace $\Pi^{\rm (V,A)} \equiv g^{\mu\nu} \Pi^{\rm (V,A)}_{\mu\nu}(q)$ contains pieces of (axial-) vectors and (pseudo-) scalars.
The pure (axial-) vector information is encoded in $\Pi^{\rm (V,A)}_{\rm T}(q)$ for which one obtains
\begin{equation}
  3  \Pi_{\rm T}^{\rm (V,A)}
    =  \frac{C^2_{\rm (V,A)}}{q^2} \Pi^{\rm (S,P)}
    + \frac{1}{q^2} \langle \bar \psi \hat q \sigma^3 \psi \rangle
        + \frac{C_{\rm (V,A)}}{q^2}
        \langle \bar \psi \tau^{\rm (V,A)} \psi \rangle
    - g^{\mu\nu} \Pi^{\rm (V,A)}_{\mu\nu}
\end{equation}
which relates $\Pi_{\rm T}^{\rm (V,A)}$ to the trace of the correlator and $\Pi^{\rm (S,P)}$.
For $m_1$ and $m_2$ being arbitrary quark masses, the chiral difference $\Pi^{\rm P-S}(q)$ and the sum $\Pi^{\rm S+P}(q)$ enter the chiral difference $q^\mu q^\nu \Pi^{\rm V-A}_{\mu\nu}(q)$ and, hence, $\Pi_{\rm T}^{\rm V-A}$.
The expectation value of the quark current cancels out in any case:
\begin{multline}
	\Pi_{\rm T}^{\rm V-A}
	= - \frac 13 \left( \frac{m_1^2+m_2^2}{q^2} \Pi^{\rm P-S}
	+ 2 \frac{m_1 m_2}{q^2} \Pi^{\rm P+S}
	\right.
	\\
	\left.
	+ \frac{2}{q^2} \left( m_1 \langle \bar{q}_2 q_2 \rangle + m_2 \langle \bar{q}_1 q_1 \rangle \right)
	+ \frac{\Pi^{\rm V-A}}{q^2}
	\right)
	\: .
\end{multline}
If one quark mass is zero, $m_1 \to 0$, one obtains the relation
\begin{equation}
\label{eq:qqva}
    q^\mu q^\nu \Pi^{\rm V-A}_{\mu\nu}(q) = - m_2^2 \Pi^{\rm P-S}(q)
    - 2 m_2 \langle \bar{q}_1 q_1 \rangle
\end{equation}
and therefore
\begin{equation}
\label{eq:vaps}
    \Pi^{\rm V-A}_{\rm T}(q) = - \frac{m_2^2}{3q^2} \Pi^{\rm P-S}(q) - \frac{1}{3} \Pi^{{\rm V-A}}(q) - \frac23 \frac{m_2}{q^2} \langle \bar{q}_1 q_1 \rangle
    \: .
\end{equation}
The last term contains the interesting combination of light-quark condensate and heavy-quark mass.

\section{Dirac projection}
\label{sct:app1}

Consider the trace over Dirac indices in $\TrD[S_2 \gamma_5 \{S_1,\gamma_5\}]$.
To write this as a product of two simpler traces, one can project the commutator
$
\{S, \gamma_5\} = \sum_{\hat{O}} A_{\hat{O}} \hat{O} 
$
with coefficients
$
A_{\hat{O}} = \frac14 \TrD[\{S, \gamma_5\} \hat{O}],
$
$\hat{O}$ being an element of the Clifford algebra. 
A similar expression applies for 
$\TrD[S_2 \gamma_5 \{\gamma_\mu S_1 \gamma^\mu,\gamma_5\}]$
with coefficients listed in Tab.~\ref{tb:app1}.
Alternatively, the coefficients for projecting 
$\Gamma^C_\mu S \Gamma_C^\mu = \sum_{\hat{O}} A_{\hat O} \hat{O}$ 
are given in Tab.~\ref{tb:app2}.

\begin{table}
\begin{tabular}{@{}ccc@{}} \toprule \hline
    $A_{\hat{O}}$           & $\{S,\gamma_5\}$                      & $\{\gamma_\mu S \gamma^\mu,\gamma_5\}$ \\ \midrule \hline
    $A_{\mathds{1}}$    & $\frac12 \TrD[S \gamma_5]$    & $-2\TrD[S]$ \\
    $A_{-}^\nu$ & $0$                                                   & $0$ \\
    $A^{\mu < \nu} $    & $\frac12 \TrD[S\gamma_5 \sigma^{\mu < \nu}]$ & $0$ \\
    $A_{+}^\nu$ & $0$                                   & $0$ \\
    $A_{5}$                     & $\frac12 \TrD[S]$                     & $2\TrD[S]$ \\ \bottomrule \hline
\end{tabular}
\caption{Coefficients of the projection of $\{S, \gamma_5\}$.}
\label{tb:app1}
\end{table}

\begin{table}
\begin{tabular}{@{}ccccc@{}} \toprule \hline
    $\Gamma^C$              & $\mathds{1}$                                                      & $\gamma_5$                                                            & $\gamma_\mu$                                                  & $\gamma_5 \gamma_\mu$ \\ \midrule \hline
    $A_{\mathds{1}}$    & $\frac14 \TrD[S]$                                             & $\frac14 \TrD[S]$                                             & $\TrD[S]$                                                         & $-\TrD[S]$\\
    $A_{-}^\nu$             & $\frac14 \TrD[S \gamma_\nu]$                      & $-\frac14 \TrD[S \gamma_\nu]$                     & $-\frac12 \TrD[S \gamma_\nu]$                 &  $-\frac12 \TrD[S \gamma_\nu]$\\
    $A^{\mu < \nu} $    & $\frac14 \TrD[S \sigma^{\mu < \nu}]$      & $\frac14 \TrD[S \sigma^{\mu < \nu}]$      & $0$                                                                       & $0$ \\
    $A_{+}^\nu$             & $\frac14 \TrD[S \gamma_5 \gamma_\mu]$     & $-\frac14 \TrD[S \gamma_5 \gamma_\mu]$    & $\frac12 \TrD[S \gamma_5 \gamma_\mu]$ & $\frac12 \TrD[S \gamma_5 \gamma_\mu]$\\
    $A_{5}$                     & $\frac14 \TrD[S \gamma_5]$                            & $\frac14 \TrD[S \gamma_5]$                            & $-\TrD[S \gamma_5]$                                   & $-\TrD[S \gamma_5]$\\ \bottomrule \hline
\end{tabular}
\caption{Coefficients of the projection of $\Gamma^C_\mu S \Gamma_C^\mu$.}
\label{tb:app2}
\end{table}

Eq.\ \eqref{eq:2} can then be obtained by using 
$\gamma_5 \sigma^{\mu\nu} = \frac i2 \epsilon_{\mu\nu\alpha\beta} \sigma^{\alpha\beta}$, 
leading to $\TrD[S_1 \gamma_5 \sigma_{\mu\nu}] 
\TrD[S_2 \gamma_5 \sigma^{\mu\nu}] = \TrD[S_1 \sigma_{\mu\nu}] \TrD[S_2 \sigma^{\mu\nu}]$.

\section{The perturbative quark propagator} \label{sct:app_prop}

The perturbative quark propagator in momentum space in a weak (classical)
gluonic background field  can be written as
$
S(p) = \sum_{n=0}^\infty S^{(n)}(p)
$
with
$
    S^{(n)}(p)
        = (-1)^n S^{(n-1)}(p) \left( \gamma \tilde{A} \right) S^{(0)}(p)
        = (-1)^n S^{(0)}(p) \left( \gamma \tilde{A} \right) S^{(n-1)}(p) \: .
$
$\tilde{A}$ denotes a derivative operator which arises due to the Fourier transform 
of the perturbation series for the quark propagator in coordinate space 
from the gluonic background field $A_{\mu}$;
$
    \tilde{A}_{\mu} = \sum_{n=0}^\infty \tilde{A}_{\mu}^{(n)}
$
with
$
    \tilde{A}_{\mu}^{(n)} = - \frac{(-i)^{n+1} g}{n! (n+2)} \left(
        D_{\alpha_1} \ldots D_{\alpha_n} \mathrsfs{G}_{\mu\nu}(0) \right)
        \partial^{\nu} \partial^{\alpha_1} \ldots \partial^{\alpha_n} .
$
The trace of each term $S^{(n)}(p)$ can be written as
\begin{equation} \label{eq:trace}
\begin{split}
    & \TrD[\Gamma S^{(n)}(p)] =
        (-1)^n \left( \frac14 \right)^n \sum_{k_1, \ldots, k_n}
        \left( \tilde{D}_{\vec{\alpha_{k_n}}} \mathrsfs{G}_{\mu_n\nu_n} \right)
        \ldots \left( \tilde{D}_{\vec{\alpha_{k_1}}} \mathrsfs{G}_{\mu_1\nu_1} \right)
        \\ & \times
        \sum_{\Gamma_1, \ldots, \Gamma_n} \TrD[\Gamma_n \Gamma S^{(0)}(p) \gamma^{\mu_n}]
        \biggl( \Bigl( \partial^{\nu_n} \partial^{\vec{\alpha_{k_n}}} \TrD[\Gamma_{n-1} \Gamma_n S^{(0)}(p) \gamma^{\mu_{n-1}}]
        \ldots \Bigr. \biggr.
        \\ & \quad \times 
        \ldots \biggl. \Bigl.
        \left( \partial^{\nu_1} \partial^{\vec{\alpha_{k_1}}} \TrD[\Gamma_{1} S^{(0)}(p)] \right)
        \Bigr) \biggr) \: .
\end{split}
\end{equation}
The sum runs over elements of the Clifford basis. 
For the sake of a concise notation we have defined
$
\tilde{D}_{\vec{\alpha_k}} \mathrsfs{G}_{\mu\nu} \equiv
        -g \frac{(-i)^{k+1}}{k!(k+2)} 
        \left( \Bigl. D_{\alpha_1} \ldots D_{\alpha_k} 
        \mathrsfs{G}_{\mu\nu} \Bigr|_{x=0} \right)
$
and
$
\partial^{\vec{\alpha_{k}}} \equiv \partial^{\alpha_1} \ldots \partial^{\alpha_k} .
$
With these reduction formulas it is only necessary to consider traces 
of the free propagator $S^{(0)} = (\hat{p}+m)/(p^2-m^2)$.
Moreover, the single traces form a chain making the sum over $\Gamma_i$ 
dependent on traces surviving from the sum over $\Gamma_{i-1}$.
In the limit $m \to 0$ and $p^2 \neq 0$ we have 
$\TrD[\Gamma_1 S^{(0)}(p, m=0)] = 0 \: \forall \Gamma \neq \gamma_\mu$. 
Hence, from the last trace in \eqref{eq:trace} only $\gamma_\mu$ is passed to the next trace.
For the latter one we therefore have 
$\TrD[\Gamma_1 \Gamma_2 S^{(0)}(p, m=0) \gamma_\mu] = 0 \: \forall \Gamma_2 \notin \{\gamma_\alpha, \gamma_5 \gamma_\alpha \}$.
As $\TrD[\gamma_5 \gamma_\alpha \Gamma_n S^{(0)}(p,m=0) \gamma_\mu] = 0 \: \forall \Gamma_n \notin \{\gamma_\alpha, \gamma_5 \gamma_\alpha \}$ also holds, each sum merely covers $\gamma_\mu$ and $\gamma_5 \gamma_\mu$.
Finally, we obtain for the first trace in the second line of \eqref{eq:trace} $\Tr[\Gamma_n \Gamma S^{(0)}(p,m=0) \gamma_\mu] = 0 \: \forall \Gamma \notin \{\gamma_\alpha, \gamma_5 \gamma_\alpha \}$, which means that $\TrD[\Gamma S^{(n)}(p,m=0)] = 0 \: \forall \Gamma \notin \{\gamma_\mu, \gamma_5 \gamma_\mu\}$.
As this is true for all orders of the perturbative sum, we can conclude that
$\TrD[\Gamma S(p, m=0)] = 0 \: \forall \, \Gamma \notin \{\gamma_\mu, \gamma_5 \gamma_\mu\}$.

\end{appendix}

\end{document}